\documentclass[]{aa}  

\usepackage{graphicx}
\usepackage{txfonts}

\usepackage{epstopdf}
\usepackage{amsmath}
\usepackage{xcolor}
\usepackage{textcomp,gensymb}
\usepackage[normalem]{ulem}
\usepackage{natbib}
\usepackage{amsmath}
\usepackage{hyperref}
\usepackage{listings}
\usepackage{color}

\definecolor{dkgreen}{rgb}{0,0.6,0}
\definecolor{gray}{rgb}{0.5,0.5,0.5}
\definecolor{mauve}{rgb}{0.58,0,0.82}
\definecolor{golden}{rgb}{0.86,0.65,0.01}
\defcitealias{Vasiliev2021}{V21}
\defcitealias{Law2010}{LM10}
\defcitealias{Penarrubia2010}{P10}
\defcitealias{Antoja2020}{A20}

\def\masyr{{\rm mas\,yr^{-1}}}
\def\HP{HEALpix }
\def\HPf{HEALpix}
\def\Lam{\tilde{\Lambda}_\odot }
\def\Bet{\tilde{\beta}_\odot }
\def\Gaia{{\it Gaia }}

\def\kms{km$\,$s$^{-1}$}
\newcommand{\FeH} {[$\mathrm{Fe}$/$\mathrm{H}$] }
\newcommand{\alphaF} {[$\alpha$/$\mathrm{Fe}$] }
\newcommand{\FeHf} {[$\mathrm{Fe}$/$\mathrm{H}$]}
\newcommand{\alphaFf} {[$\alpha$/$\mathrm{Fe}$]}

\begin{document}

   \title{The Sagittarius stream in Gaia eDR3\\ and the origin of the bifurcations}

   %\titlerunning{}

   \author{P. Ramos\inst{1}\fnmsep\thanks{email: p.ramos@unistra.fr}
            \and
          T. Antoja\inst{2}
            \and
        Z. Yuan\inst{1}
                \and
        A. Arentsen\inst{1}
                \and
        P.-A. Oria\inst{1}
                \and
         B. Famaey\inst{1}
                \and
        R. Ibata\inst{1}
                \and
          C. Mateu\inst{3}
            \and
          J.A. Carballo-Bello\inst{4}}

   \institute{Observatoire astronomique de Strasbourg, Universit{\'e} de Strasbourg, CNRS, 11 rue de l’Universit{\'e}, 67000 Strasbourg, France
        \and
            Dept. FQA, Institut de Ci{\`e}ncies del Cosmos (ICCUB), Universitat de Barcelona (IEEC-UB), Mart{\'i} Franqu{\`e}s 1, E-08028 Barcelona, Spain
         \and
             Departamento de Astronom{\'i}a, Instituto de F{\'i}sica, Universidad de la Rep{\'u}blica, Igu{\'a} 4225, CP 11400 Montevideo, Uruguay
        \and
            Instituto de Alta Investigaci{\'o}n, Sede Esmeralda, Universidad de Tarapac{\'a}, Av. Luis Emilio Recabarren 2477, Iquique, Chile}
   \date{Received XXX; accepted XXX}

% \abstract{}{}{}{}{}  
% 5 {} token are mandatory
 
  \abstract
   {The Sagittarius dwarf spheroidal (Sgr) is a dissolving galaxy being tidally disrupted by the Milky Way (MW). Its stellar stream still poses serious modelling challenges, which hinders our ability to use it effectively as a prospective probe of the MW gravitational potential at large radii.} 
   {Our goal is to construct the largest and most stringent sample of stars in the stream with which we can advance our understanding of the Sgr-MW interaction, focusing on the characterisation of the bifurcations.}
   {We improve on previous methods based on the use of the wavelet transform to systematically search for the kinematic signature of the Sgr stream throughout the whole sky in the \textit{Gaia} data. We then refine our selection via the use of a clustering algorithm on the statistical properties of the colour-magnitude diagrams.}
   {Our final sample contains more than 700k candidate stars, 3x larger than previous \textit{Gaia} samples. With it, we have been able to detect the bifurcation of the stream in both the northern and southern hemispheres, requiring four branches (two bright and two faint) to fully describe this system. We present the detailed proper motion distribution of the trailing arm as a function of the angular coordinate along the stream showing, for the first time, the presence of a sharp edge (on the side of the small proper motions) beyond which there are no Sgr stars. We also characterise the correlation between kinematics and distance. Finally, the chemical analysis of our sample shows a significant difference between the faint and bright branches of the bifurcation. We provide analytical descriptions for the proper motion trends as well as for the sky distribution of the four branches of the stream.}
   {Based on our analysis, we interpret the bifurcations as the misaligned overlap of the material stripped at the antepenultimate pericentre (faint branches) with the stars ejected at the penultimate pericentre (bright branch), given that Sgr just went through its perigalacticon. The source of this misalignment is still unknown but we argue that models with some internal rotation in the progenitor -- at least during the time of stripping of the stars that are now in the faint branches -- are worth exploring.}


   \maketitle
%
%-------------------------------------------------------------------

\section{Introduction}\label{sec:intro}

The Sagittarius dwarf galaxy (Sgr) was discovered serendipitously more than 25 years ago by \citet{Ibata1994,Ibata1995}. Since then, it has sparked the interest of many astronomers for it is the closest dwarf galaxy that we can study and for the ability of its tidal stream to constrain the gravitational potential of the Milky Way (MW). The dwarf galaxy is composed of a peculiar mix of old, intermediate-age and young populations of stars \citep[e.g.,][]{Sarajedini1995,Layden2000,Siegel2007,deBoer2014,Hasselquist2021}, the latter probably formed during the last disc crossing \citep{TepperGarcia2018}. It was clear from the beginning that it is undergoing full tidal disruption, and it did not take long until the first hints of the tidal tails were found \citep{Mateo1996,Alard1996,Fahlman1996,Mateo1998}. But it was not until the arrival of all-sky photometric surveys that the whole extent of its stellar stream was revealed \citep{Ibata2001,Majewski2003}. After that, many works have re-detected the stream with newer and better data; the latest samples \citep[e.g.,][]{Antoja2020,Ibata2020,Ramos2020} coming entirely from \Gaia data \citep{gaiamission}. The stellar stream, an almost polar structure of tidally stripped material, is divided into two arms: the leading (most prominent in north galactic hemisphere, goes ahead of the progenitor since it is at inner Galactic radii) and the trailing (most prominent in south galactic hemisphere, trails behind the progenitor since it is at outer Galactic radii) -- as expected from a dissolving stellar system. However, the picture became much more complex after the discovery of secondary branches, usually referred to as bifurcations, in both the leading \citep{Belokurov2006} and trailing \citep{Koposov2012} tails. 

Early attempts at modelling Sgr, given the available data at the time, focused on reproducing the current state of the remnant \citep[e.g.,][]{Velazquez1995,Johnston1995}. Interestingly, \citet{Ibata1997} even explored the possibility that the dwarf hosted a rotating disc, concluding, based on their simulations, that it is unlikely given the observations \citep[as latter shown by][]{Penarrubia2011}. Once the stream was discovered, the models shifted their focus onto reproducing the stream, which quickly led to contradicting results with regards to the shape of the dark matter halo. While the radial velocity trends of the leading arm seem to require a prolate halo \citep{Helmi2004}, the difference in the mean orbital poles between the two tails favours an oblate halo instead \citep{Johnston2005}. \citet[][hereafter LM10]{Law2010} solved this tension by requiring a triaxial halo. Notwithstanding, their resulting mass distribution has its minor axis oriented along the Galactic plane, in principle an unstable configuration, which led the authors to conclude that other non-axisymetric effects like, for instance, the influence of the Magellanic Clouds should be taken into account. In spite of that, later models, like those of \citet{Gibbons2014}, \citet{Dierickx2017} or \citet{Fardal2019}, use mostly spherical models with different radial profiles. Recently, though, \citet[][hereafter V21]{Vasiliev2021} proposed instead a twisted halo that transitions from prolate in the outer parts to oblate in the inner parts \citep[see][although they find it oblate in the outer parts]{Shao2021} which, once combined with the effect of the infalling massive Large Magellanic Cloud (LMC), produces an excellent agreement with the \textit{Gaia} second data release (DR2). It is important to point out, however, that all these models are based mostly on fits to the younger part of the stream (dominated by the material stripped at the penultimate pericentre), which has less constraining power than the material stripped at the antepenultimate pericentre (i.e., $\gtrsim$2Gyr ago), and completely neglect the presence of the bifurcations whose origins remain a mystery.

As it can be seen, there has been a lot of effort devoted equally to understanding Sgr and inferring MW properties from it. At this point, one of the biggest challenges remaining is the formation of the aforementioned bifurcations. In this work, we compile the largest sample to date of stars in the Sgr stream with the methods described in Sect.~\ref{sec:data&methods}. We then present the main properties of this sample and the new constraints it will allow in Sect.~\ref{sec:results}. More importantly, Sect.~\ref{sec:discussion} is dedicated to a re-evaluation of the nature of the bifurcation in the light of the new data and its possible origin. As we conclude in Sect.~\ref{sec:conclusions}, we find that the most plausible explanation for the bifurcations is that their faint branches are made of stars stripped shortly 'after' the antepenultimate pericentre passage from a Sgr dwarf galaxy that either had some internal rotation or that suffered a perturbation on its way to the next pericentre, ejecting material with slightly different orbital properties.

%--------------------------------------------------------------------
\section{Data and methods}\label{sec:data&methods}

\subsection{Data}\label{sec:data}
The early third data release of the \Gaia catalogue \citep[eDR3][]{Brown2021} contains astrometric solutions for roughly 1.5 billion sources among which our target, the Sgr stream, lies. Although the selection function of \Gaia is, for the moment, not known with precision \citep[but see][]{Everall2021}, we do know that many of the observed stars are close to the Sun, blocking our view of the halo, the outer disc and, of course, the Sgr stream. In an attempt to reduce the foreground contamination, in \citet{Antoja2020} we adopted a simple cut in parallax, $\varpi-\sigma_\varpi < 0.1\,$mas, which by construction preserves most of the stars farther than 10\,kpc from the Sun while filtering most of the nearby stars. However, to first order, the parallax distribution of Sgr stars is a Gaussian centred at zero whose dispersion is dominated by the formal errors. As a result, this cut removes a significant part of the stars in the stream (roughly, the positive 2-sigma tail) and introduces obvious biases in the distribution of parallaxes which invalidates any attempt to obtain valuable information from this observable.

In this work, instead, we used the following cut:
\begin{equation}\label{eq:plx_cut}
    |\frac{\varpi}{\sigma_\varpi}| < 4.5,
\end{equation}
\noindent which can be understood as selecting only the stars with poor parallaxes. This filter does not bias the distribution of parallaxes for distant systems such as the Sgr stream and removes most of the foreground quite efficiently either because i) the source has a small parallax uncertainty or ii) the parallax is large. The exact value of 4.5 is motivated by the work of \citet{Rybizki2021} but, after checking the particular case of Sgr, we note that setting the value to $\sim3$ could have been enough which is, incidentally, the value below which parallaxes are hardly informative \citep[see Appendix C.2 of][]{Antoja2021}.

Another improvement with respect to our previous works is that now we removed the quasars from our sample upfront using the table {\normalfont\ttfamily{agn\_cross\_id}} provided with the eDR3 catalogue \citep{Lindegren2021}. While there might still be some quasars left, now their density should be low enough to have a negligible contribution \citep[see][for a description of the impact that quasars have on the search for kinematic substructure in the halo]{Ramos2021}.

The resulting sample contains 1\,248\,862\,405 stars. We further constrained the sample to the plane of the Sgr orbit: $|\Bet|\,<\,$25$^\circ$, where $\Lam$ and $\Bet$ are, respectively, the spherical angular coordinates along and across the Sgr stream (first introduced by \citealt{Majewski2003}; later \citealt{Belokurov2014} defined the convention that we used in this work). We also avoided the MW disc by removing sources in the set $(\Lam$<-150$^\circ) \cup (\Lam$>160$^\circ) \cup (\Lam$>10$^\circ \cap \Lam$<30$^\circ$). After these final cuts, we reduced our sample to 238\,687\,820 sources.

\subsection{Methodology}\label{sec:methods}

\begin{figure*}[]
   \centering
    \includegraphics[width=0.98\textwidth]{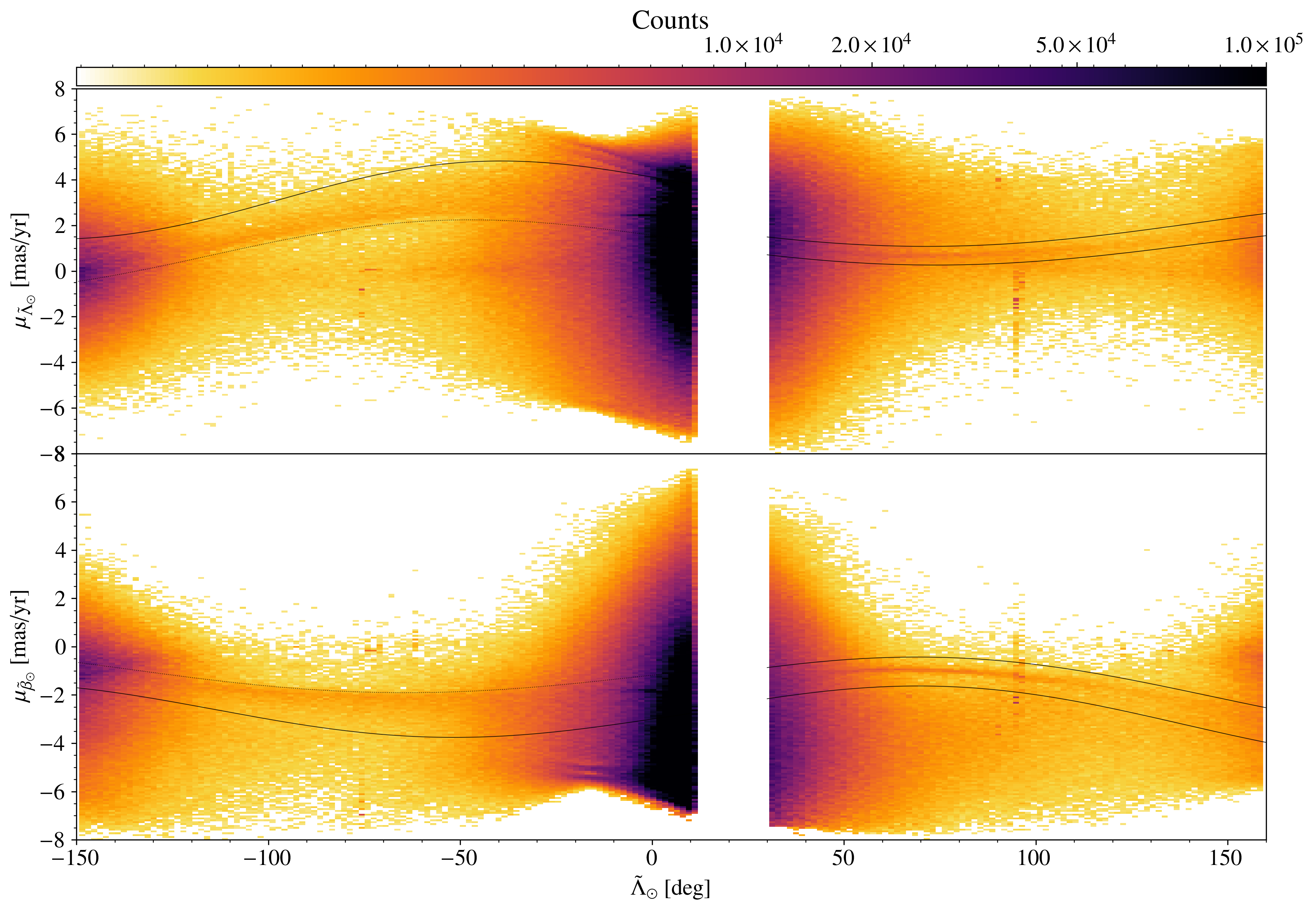}
    \caption{Proper motion of the kinematic trees in the Sgr celestial frame as a function of $\Lam$. Top: $\mu_{\Lam}$ coordinate of the kinematic structures detected, weighted by the number of stars in each tree (scale of maximum WT coefficient). Bottom: same, but in $\mu_{\Bet}$. The footprint of the Sgr stream is clearly distinguishable in both the leading and trailing arms. In the latter, we note a sharp edge and a diffuse component above/below it (top/bottom respectively). The lines represent the upper and lower limits of our kinematic selection.}
    \label{fig:fig1} 
\end{figure*}

The data was analysed in a similar fashion to \citet{Antoja2020} and \citet{Ramos2021} with a few, but important, improvements. Overall, the goal of our technique is to detect kinematic substructures -- those related to the Sgr stream in particular-- in the proper motion histograms throughout our whole sample by analysing in parallel each \HP bin in the sky. The main steps of this methodology can be summarised as follows (see Appendix~\ref{app:methods} for a more detailed description): use the Wavelet Transformation \citep[WT,][]{Starck2002} to decompose a proper motion histogram into layers, each containing structures of similar sizes, and find all the significant kinematic over-densities. Then, use the information of the stars that contributed to each over-density to find out its nature (MW component, globular cluster, dwarf galaxy, etc.).

The first improvement comes from increasing the accuracy and sharpness of our detection. We did so by downloading the proper motion histograms (bin size of 0.12 mas\,yr$^{-1}$) for every \HP level 6 tile (roughly 0.84 square degrees each) in our $|\Bet|\,<\,$25$^\circ$ footprint, using the query \ref{q1} in Appendix \ref{app:queries}. This represents a reduction by half in the bin size of the proper motion histograms and an increase by one of the \HP level with respect to our previous works.

The second improvement corresponds to the way we treated the kinematic substructures detected with the WT at each of the proper motion histograms downloaded. By construction, the WT of every histogram yields dozens of over-densities (peaks), and not all of them are relevant to us. In the past, we had to ignore all but one peak per \HP in order to handle the large amount of information. This time, instead of considering the over-densities found at different layers of the WT as independent, we grouped them hierarchically in what we call 'kinematic trees'. In other words, we try to associate together all peaks that belong to a single stellar object across all WT layers. These trees completely characterise one kinematic structure with their ra-dec-pmra-pmdec coordinates and their size in proper motion space.

The third main improvement refers to the size of these structures. By comparing the \citet{Antoja2020} and \citet{Ibata2020} samples we noticed that, in the former, we used a radius around the WT peak too small, reducing significantly the completeness of our sample. Thus, whenever we needed to select stars from a kinematic tree we used four\footnote{This value is arbitrary and has been chosen after experimenting with other sizes.} times its characteristic size.

\begin{figure*}[]
   \centering
    \includegraphics[width=0.98\textwidth]{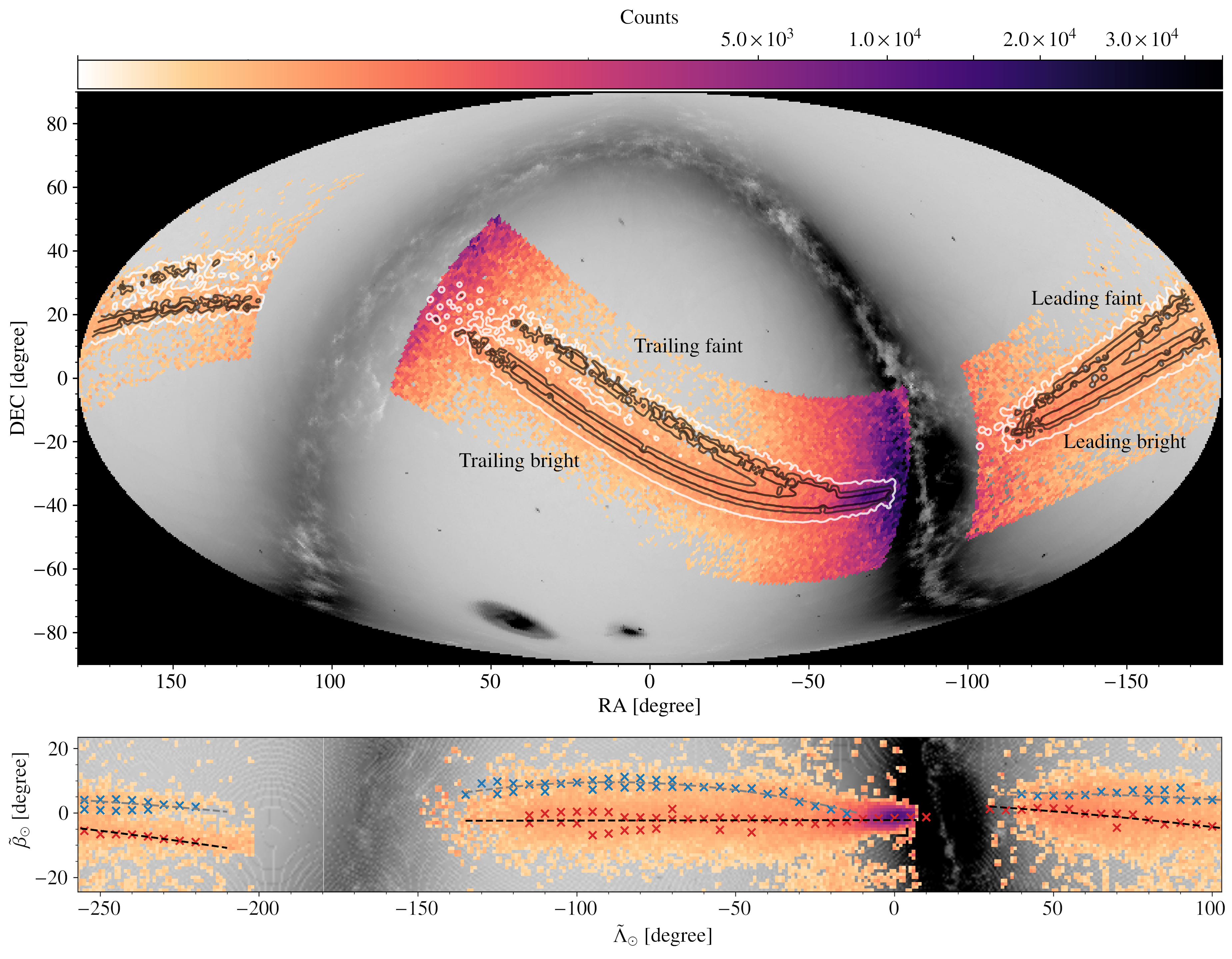}
    \caption{Sky distribution of the candidate Sgr stream stars. Top: histogram, in ICRS coordinates, of the counts obtained from the kinematic trees selected in proper motion (Sect.~\ref{sec:methods}). The contours are iso-probability lines of Prob(Sgr|$\Lam, \Bet$), a score assigned to each source in the final catalogue based on the probability of being a Sgr stream star (see text). On the background, we included a black and white histogram of the input sample for reference. Bottom: histogram of the stars in our final sample, with bins of 1$^\circ$x1$^\circ$, but in the Sgr reference frame ($\Lam$, $\Bet$). The red (blue) crosses correspond to the peaks of the 1D WT associated with the bright (faint) branch, while the dashed lines represent the resulting polynomials (see Table~\ref{tab:bifu_polys}). The Sgr stream appears as a four-tailed structure across the whole sky, for the first time, in a homogeneous astrometric sample.}
    \label{fig:fig2} 
\end{figure*}

Once we ran our method on the sample described in the previous section, we obtained a list of 1 207 419 kinematic trees. The large majority of them do not correspond to our target, the Sgr stream. To select only the structures that we are interested in we first looked for the signal of the stream in proper motion. Figure \ref{fig:fig1} shows the sum of the stellar counts of all the kinematic trees in the space of $\mu_{\Lam}$ and $\mu_{\Bet}$ versus $\Lam$. The first thing we noted is the small over-densities, some of them in the form of a vertical stripe, which are caused by globular clusters (e.g., the feature at $\Lam \sim 95^\circ$ is a combination of the kinematic signatures of M3, M53 and NGC5053). Apart from that, we see the prominent over-density caused by the thick disc at $-40^\circ \leq \Lam \leq 50^\circ$ and that the halo stars create a horizontal feature at  $\mu_{\Lam} \sim 0\,\masyr$. 

Amidst these two populations, the signal of the Sgr stream appears clearly in what at first sight seems to be a thin kinematic structure. Upon closer inspection, we realised that the trailing arm of the stream ($\Lam < 0^\circ$) is better described by a sharp edge, marked by the dotted lines in Fig.~\ref{fig:fig1}, and a diffuse envelope (solid black lines). In both $\mu_{\Lam}$ and $\mu_{\Bet}$, we observe a decrease in density going from the dotted line to the solid line. To the best of our knowledge, this is the first time we can reach this level of detail with observational data. In the leading arm ($\Lam > 0^\circ$) we do not note any substructure, probably due to the fact that this portion of the stream is mostly at larger heliocentric distances. We then adapted the shape proposed by \citet{Ibata2020}, and reproduced in Eq.~\ref{eq:muVSlambda} for convenience, to describe the shape of the proper motion trends of the Sgr stream. The result is the 8 lines shown in Fig.~\ref{fig:fig1}, for which we used the parameters of Table~\ref{tab:parameters}. Consequently, we only selected the kinematic trees that fall inside the contours delineated by said lines.

\begin{equation}\label{eq:muVSlambda}
\mu_{X}(\Lam) = a_1\textrm{sin}(a_2\Lam + a_3) +a_4 +a_5\Lam +a_6\Lam^2    
\end{equation}

The resulting selection of peaks for the Sgr stream can be seen in Fig.~\ref{fig:fig2} with the histogram coloured by counts. It is clear that there is still a significant fraction of contamination from both the thick disc and halo components. This can also be confirmed when looking at the colour-magnitude diagram (CMD) of the stars contributing to each individual kinematic tree, since halo and thick disc stars have CMDs clearly distinct from those of the Sgr stream. The most telling factor is the sign of the correlation between colour and apparent magnitude: Sgr presents the typical shape of a red giant branch (negative correlation), whereas the contamination has a triangle shape with a positive correlation (see Fig.~\ref{fig:figB1}).

We introduce a novel technique to increase the purity of the sample based on the distinct CMD track that each population has. First, we used query~\ref{q2} in Appendix \ref{app:queries} to count the number of stars that each kinematic tree contains along with four indicators that summarise the main properties of their CMDs: mean colour $G_{BP}-G_{RP}$, mean magnitude $G$, spread in colour $\sigma_{G_{BP}-G_{RP}}$ and, finally, the Pearson correlation $r_{G-colour}$. Then, we applied a simple k-means clustering algorithm to the summary statistics of the CMDs that we obtained for each peak, in combination with their respective average $\Lam$ and $\Bet$, after properly normalising the different quantities. We used 6 components to represent the three main populations present in our data: Sgr, halo and disc. Each of them separate cleanly even with such a simple set-up. This produces a list of kinematic trees candidates that potentially belong to the Sgr stream.

After this last step we obtained the list of candidate substructures, from which we extracted the final sample of 774\,374 candidate stars brighter than $G\,<\,19.75\,\textrm{mag}$, shown in Table~\ref{tab:sample} and available online.  Compared to previous samples, this sample is almost three times as big and, in particular, it contains $\sim$75\% of the 294 344 sources in \citet{Antoja2020} and $\sim$72\% of the 263 438 sources in \citet{Ibata2020}. While it is tempting to classify all the missing sources in those catalogues as contaminants, it is worth mentioning that, due to some of the steps introduced in our method, the completeness in some regions of the sky is not as high as we would wish. Nonetheless, our sample contains 8 084 RR-Lyrae, which is almost exactly the value predicted by \citet{Cseresnjes2001b} and right in between the two samples (one pure but incomplete, and one complete but contaminated) given in \citet{Ramos2020}. 

We complemented our sample with radial velocities obtained from the  Apache Point Observatory Galactic Evolution Experiment \citep[APOGEE,][]{Majewski2017} and the Sloan Extension for Galactic Understanding and Exploration \citep[SEGUE,][]{Yanny2009}, both obtained from the Sloan Digital Sky Survey DR16 \citep{Ahumada2020}, the Large Sky Area Multi-Object Fiber Spectroscopic Telescope \citep[LAMOST DR6,][]{Cui2012,Zhao2012}, Gaia DR2 \citep{Brown2018}, and SIMBAD. We combined all radial velocities (after some minor quality filters, see Appendix~\ref{app:methods_vlos}) into a single value for each source by taking their median. We did the same for the spectroscopic metallicities but, this time, we did not merge the different catalogues into one value. Instead, we treated each catalogue independently. 
Finally, we also included reddening-free Wesenheit distances (computed from the \Gaia DR2 $G$ band and BP-RP colours) for our sub-sample of RR-Lyrae using the calibration of \citet{Neeley2019}.

%--------------------------------------------------------------------
\section{Results}\label{sec:results}

\begin{figure*}[]
   \centering
    \includegraphics[width=0.98\textwidth]{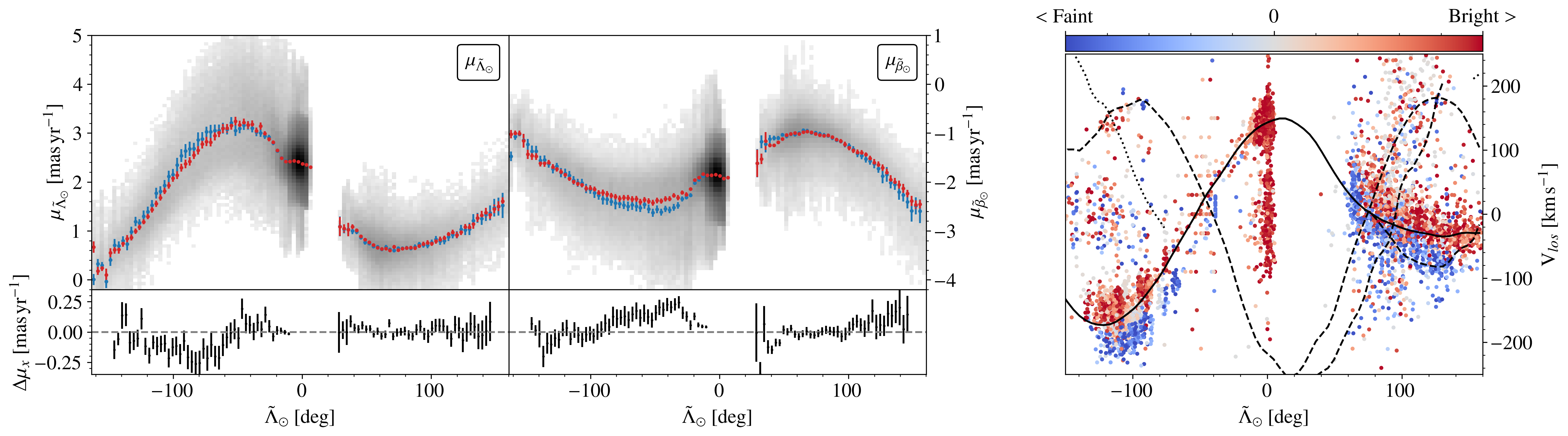}
    \caption{Proper motions and radial velocities of the bright (red) and faint (blue) branches of the Sgr stream. Left: proper motions of the two branches and their differences with 3$\sigma$ error bars. Right: radial velocity coloured by $\Bet$ for the sources with Prob($\varpi$) > 0.8 (see text). The solid/dash/dotted lines correspond to the first/second/third wraps of the \citetalias{Vasiliev2021} model. Although the differences between the bright and faint branches are significant, these are most likely caused exclusively by the correlation between distance, proper motion and $\Bet$.} 
    \label{fig:fig3} 
\end{figure*}

In this section, we analysed the changes of each dimension (position, kinematics, chemistry), and their correlations, along the whole stream.

\subsection{Spatial distribution}
Already from a visual inspection of the sky density of our sample (see bottom panel of Fig.~\ref{fig:fig2}) we identified two over-densities in the sky running parallel to the main spine of the stream, which clearly correspond to the faint branches of the bifurcation: the leading in the northern hemisphere \citep{Belokurov2006} and trailing in the southern hemisphere \citep{Koposov2012}. Compared to the main, brighter stream, these other two branches appear fainter in the sky mostly because they contain less stars. We note that we cannot observe a significant signal for the faint branch in the trailing arm beyond $\Lam \gtrsim -40^\circ$. While it is tempting to associate the loss of a secondary branch with the increase in foreground contamination, this is unlikely to be the case since the RR Lyrae sample is almost complete and allows us to track the stream into much denser regions. And yet, we did not observe any evidence for a secondary stream at $\Lam > -40^\circ$ neither with the RR Lyrae sub-sample in this work nor in \citet{Hernitschek2017} or \citet{Ramos2020}. In consequence, it seems more likely that either the stars in the faint branch fall outside of our detection limits (proper motion and/or magnitude) beyond this range in $\Lam$ or that the bifurcation simply ends, meaning that its angular separation from the dwarf is in fact physical. However, as noted in \citet{Navarrete2017}, this could also be due to the lack of observable RR Lyrae stars so close to the Galactic plane. Also, \citet{Navarrete2017} looked for the bifurcation with other tracers and detected it almost to the edge of the footprint of the photometric survey they used ($\Lam \sim -40^\circ$).

\begin{table}
\caption{Coefficients of the second order polynomials used to describe each of the four arms of Sgr.}\label{tab:bifu_polys}
\centering
\begin{tabular}{cccc}
\hline\hline
  Arm/Branch   & a & b  & c      \\
\hline
Trail/Bright & 0.000 & 1.654$\times 10^{-3}$ & -2.214 \\
Trail/Faint & -1.563$\times 10^{-3}$ & -2.805$\times 10^{-1}$ & -3.040 \\
Lead/Bright & -3.260$\times 10^{-4}$ & -4.828$\times 10^{-2}$ &  3.708 \\
Lead/Faint & -3.819$\times 10^{-4}$ & 1.904$\times 10^{-2}$ &  6.084 \\
\hline
\hline           
\end{tabular}
\tablefoot{Each polynomial is of the form $\Bet(\Lam) = a\Lam^2 + b\Lam + c$, with both angular coordinates expressed in degrees.}
\end{table}

To better quantify the shape of the branches, we ran a 1D WT on the smoothed $\Bet$ histograms of the kinematic trees (weighted by counts) every 5$^\circ \pm$10$^\circ$ in $\Lam$. From the WT obtained at each bin in $\Lam$ we extracted the $\Bet$ values of the peaks. Some of these peaks should correspond to the centre of every over-density present in the $\Bet$ histograms. And, indeed, we found that the highest ones trace accurately the spine of the bright branch -- the main branch-- along the whole stream. We also detected a coherent trace of peaks (usually the second or third in height) that follow the over-densities observed by eye and corresponding to the faint branches. Therefore, we assigned each peak accordingly to either the faint or bright branch, from which we obtained four sequences in $\Lam$-$\Bet$ coordinates (two for the leading arm and two for the trailing). Finally, we fitted second order polynomials of the form $\Bet(\Lam) = a\Lam^2 + b\Lam + c$ to the sky coordinates of each of them, obtaining the coefficients reported in Table~\ref{tab:bifu_polys}. To then confirm our results, first we ran a suite of Bayessian Gaussian Mixture models, which resulted in a similar description of the sky distribution of the stream. However, we preferred to use the 1D WT since the only assumption it requires is that the peak of the WT coincides with the peak of the density, which we tested with some toy models built from a simple superposition of Gaussians with different means and dispersion. Then, we compared our results to the $\Bet$ histograms of the RR Lyrae sub-sample, obtaining a good match and also compatible with the results of \citet[][leading arm only]{Ramos2020}.

Having obtained a mathematical description of the branches, we would also like to quantify the probability of any star in our sample of belonging to any of the four branches of Sgr. To do so, however, we must assume a width along the tails. For simplicity, we chose to model each branch as Gaussians of constant width. In contrast to the fits of \citet{Koposov2012}, we chose smaller widths, $\sigma_{Bright} = 2.5^\circ$ and $\sigma_{Faint} = 1.5^\circ$, to describe our data. With this, we obtained a simple way of separating Sgr's four arms and, also, a way of measuring the probability of any given star in our final sample of belonging to the stream based on its sky position with 

\begin{equation}\label{eq:prob_sgr}
\begin{aligned}
    \textrm{Prob(Sgr}\,|\,\Lam, \Bet) = \textrm{Prob(A}\,|\,\Lam, \Bet) + \textrm{Prob(B}\,|\,\Lam, \Bet)\\ - \textrm{Prob(A}\,|\,\Lam, \Bet)\textrm{Prob(B}\,|\,\Lam, \Bet),
\end{aligned}
\end{equation}

\noindent where Prob(A|$\Lam, \Bet$) and Prob(B|$\Lam, \Bet$) are the normalised (the mode has probability of 1) Gaussian probabilities of belonging, respectively, to the bright and faint branches. In Fig.~\ref{fig:fig2} we show the contours delineating three levels of the probability Prob(Sgr|$\Lam, \Bet$) of our final sample (black and white lines). For the first time in an all sky astrometric sample of individual stars, the four arms can be clearly appreciated.

\subsection{Kinematics}
\begin{figure*}
   \centering
    \includegraphics[width=0.95\textwidth]{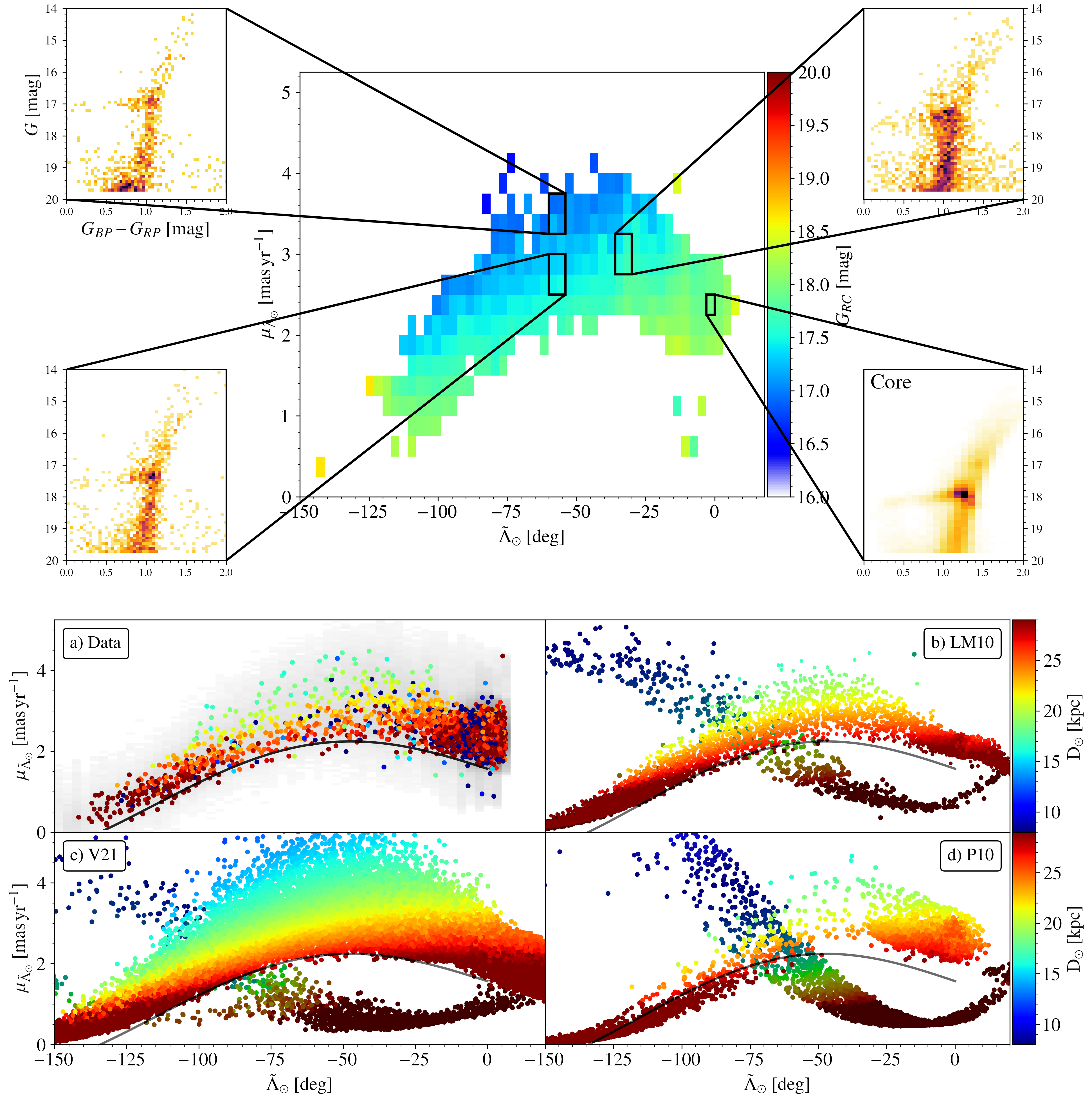}
    \caption{Tangential proper motion as a function of $\Lam$, coloured by different distance estimators, for the trailing arm. Top: coloured by the G magnitude of the red clump, as exemplified by the CMDs surrounding the panel (no extinction correction). Bottom: same projection, now coloured by distance, but for the RR Lyrae in our sample (a) as well as 3 different N-body simulations (b for \citetalias{Law2010}, c for \citetalias{Vasiliev2021} and d for \citetalias{Penarrubia2010}). We note a clear correlation between the modulus of the proper motion vector and the distance, with small distances corresponding to larger proper motions. The data is, in general, well reproduced by the recently stripped material (light colours). The black lines correspond to the lower limit used in our selection of the kinematic trees (see Table~\ref{tab:parameters}). The \citetalias{Vasiliev2021} is the model that better represents the data. In darker colours, we show the second wrap of the leading arm and we note that each model predicts a different distribution and region of overlap with the trailing tail.}
    \label{fig:fig4} 
\end{figure*}

Next, we used the mathematical description of each arm to study the kinematics of the stream. After weighting the proper motions and radial velocities with either $\textrm{Prob(A}\,|\,\Lam, \Bet)$ or $\textrm{Prob(B}\,|\,\Lam, \Bet)$ we noted that the differences between the bright and faint branches, despite being significant, can be easily attributed to projection effects (see Fig.~\ref{fig:fig3}). For instance, in the case of the proper motions, we observe a similar trend with $\Lam$ when comparing the centre of the stream with the stars at the symmetric location of the faint branch with respect to the bright branch. In the case of the radial velocities, there is a smooth transition from positive to negative $\Bet$, with the physical centre of the stream lying in the middle of the radial velocity track. We have also looked at the distances (using both RR Lyrae and the median apparent magnitude) and did not find any evidence of a bi-modality. In other words, we did not observe the existence of two populations in any phase-space dimension other than in the sky position.

Figure \ref{fig:fig3}, right panel, also includes the predictions of the \citetalias{Vasiliev2021} N-body model for the radial velocities as a function of $\Lam$. As can be seen, our sample contains not only the main branches (solid lines), some halo and thick disc contamination (cloud of points centred around zero radial velocity), but also older wraps as well (dashed and dotted lines). Indeed, our results are consistent with previous 6D samples such as \citet{Yang2019} and \citet{Penarrubia2021}. In the latter, the authors report also the detection of old material in the trailing arm. In our sample, this corresponds to the diffuse cloud of points at $\Lam\sim100^\circ$. Analogously, we associated the small clump of stars at $\Lam\sim-100^\circ$ and V$_{los}\sim150\,$\kms{} with the leading arm. Interestingly, though, we only detected the older leading arms at the point where the second and third wraps should cross each other. The use of radial velocities would allow us to, on one hand, obtain a purer selection of 6D stars and, on the other, to constrain the past orbit of Sgr with much better accuracy. 
 
 Focusing now on the trailing arm, which we can analyse in greater detail since it is much closer to us than most of the leading, in Figure~\ref{fig:fig4} we studied the correlations between the proper motions, in particular $\mu_{\Lam}$, and the distances. For this exercise we did not make any distinction between the bright and faint branches. The advantage of this sample, being so large, is that we can obtain detailed CMDs for each portion of the $\mu_{\Lam}$ - $\Lam$ diagram (or $\mu_{\Bet}$ - $\Lam$ for that matter) as exemplified by the small histograms surrounding the top panel of Fig.~\ref{fig:fig4}. 
 Then, we set out to measure the peak in the apparent magnitude histogram, corresponding to the $G$ magnitude of the Red Clump, $G_{RC}$, and use that as tracer of the distance. To this end, we first fitted a Gaussian kernel to the $G$ histogram of the sources in each bin in $\Lam$ vs $\mu_{\Lam}$, from which we then extracted the peaks -- using the same code we use for the WT peaks-- and sort them by height. The peak of the RC is usually the highest one. In some cases, though, the highest peak corresponds to the magnitude limit imposed (i.e., the histogram diverges towards the faint end due to the presence of the main sequence). When this happens, we simply took the second highest peak. As a result, we coloured each $\Lam$ vs $\mu_{\Lam}$ bin by the corresponding $G_{RC}$, as can be seen in the top panel of Fig.~\ref{fig:fig4}, from which it is obvious that there is a correlation between the modulus of the proper motion vector and the heliocentric distance. This tendency is further confirmed using RR Lyrae distances\footnote{Here, we have used a simple Period-Wesenheit relation \citep{Neeley2019} and, as result, the distances may be suffering from biases since we did not consider the metallicity. However, the trends observed cannot be caused by the lack of metallicities in the distance calculation since said metallicites would have to differ by more than $\sim$2~dex along $\mu_{\Lam}$ in order to account for the gradient, which we do not observe.}, shown in panel (a) of the same figure. 

 While this correlation is, to the best of our knowledge, the first time that it has been reported observationally, it is somewhat expected. The physical explanation is simple: stars on the stream sharing the same 3D velocity will have larger proper motions if they are closer to the observer. Similarly, the stars that are farther will pile up into an over-density at small proper motions, thus creating the sharp edge that we parametrised in Table~\ref{tab:parameters}. This limit is set by the outer layer of the stream, that is, its far-side envelope. Therefore, it tells us about the morphology of the stream in configuration space. What is less intuitive, and yet also completely expected, is that there is an underlying correlation between the distance and the position across the stream, $\Bet$ (not shown here). This latter correlation arises due to the fact that the stream appears broader on the sky when considering the nearby stars, and more collimated when taking only the stars farther away from us. In other words, the stream seems to "fan out" in the sky as the $\Bet$ distribution of the stellar debris grows broader with decreasing heliocentric distance. It is therefore more likely to have a star close to us at large values of $\Bet$, and therefore also at large values of $|\mu_{\Lam}|$ and $|\mu_{\Bet}|$.

 In this regard, we did not see any evidence of two different populations at high $\Bet$ neither in proper motion space (which is obvious since we forced a cut in proper motion space) nor in distance\footnote{This seems to contradict the results of \citet{Slater2013}, who finds the faint branch significantly closer to the Sun than the bright arm. However, as shown in \citet{Navarrete2017}, their detection is not actually related to the faint branch as we defined it in this work.} (only one Red Clump at any given position on the sky). This in turn means that, whatever the nature of the bifurcation is, it must behave very similar to the canonical first wrap of the trailing arm within the range $-120^\circ \lesssim \Lam \lesssim -50^\circ$. 
 
Panels (b), (c) and (d) of Fig.~\ref{fig:fig4} show the same space as (a) but for the stellar particles obtained from three different N-body models created to replicate the Sgr stream. Respectively, these are the \citetalias{Law2010}, \citetalias{Vasiliev2021} and \citet[][P10]{Penarrubia2010} models. The difference between \citetalias{Law2010} and \citetalias{Penarrubia2010} is just the internal dynamics of the progenitor as, in case of the latter, Sgr is a rotationally supported system. On the other hand, the main difference between \citetalias{Law2010} and \citetalias{Vasiliev2021} is the inclusion of the LMC, which in turn also requires modifying the shape of the MW halo to produce a stream compatible with Sgr. All three models predict a similar first wrap for the trailing arm (in all projections of phase-space), but the \citetalias{Vasiliev2021} model is the one that best reproduces the observations, as can be seen by comparing the lower envelope of their proper motion trends with the lower bound that we obtained from the data (black solid line). 
The biggest difference between the models, though, is in their ancient stripped material (darker colour map), as each model predicts a different trend and site of crossing with the trailing arm. Due to the potentially high constraining power of this old branch, it is very valuable. 

 \begin{figure}[h!]
   \centering
    \includegraphics[width=0.455\textwidth]{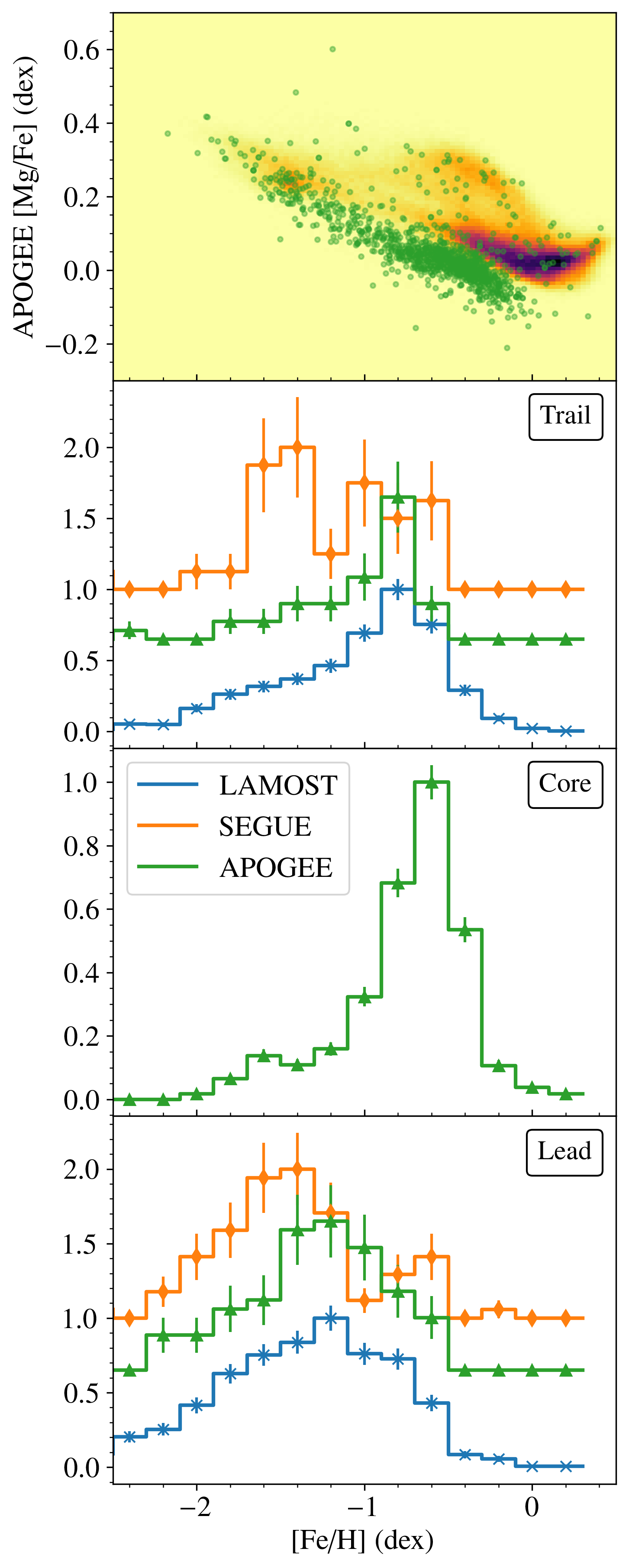}
    \caption{Chemical composition of the Sgr stream. Top panel: \alphaF against \FeH of the stream with APOGEE data (in the background, the same but for the whole APOGEE sample). Bottom panel: normalised metallicity distribution function for the trailing, core and leading parts (top, middle and bottom respectively) in bins of 0.2~dex, shifted vertically for clarity. All three LAMOST, APOGEE and SEGUE samples agree that the leading arm is more metal poor than the trailing.}
    \label{fig:fig5} 
\end{figure}

\subsection{Chemistry}

\begin{table}[t]
\caption{Mean metallicity (\FeHf, in dex, top) and mean alpha over iron (\alphaFf, in dex, bottom) of the different portions of the Sgr stream.}\label{tab:metallicites}
\centering
\begin{tabular}{cccc}
\hline\hline
    & APOGEE & LAMOST  & SEGUE\\
\hline
\FeHf & & &\\
Trail & -0.85$\pm$0.06             & -0.84$\pm$0.02 & -1.12$\pm$0.01 \\
Lead & -1.12$\pm$0.04              & -1.15$\pm$0.02 & -1.41$\pm$0.02 \\
Lead/bright & -1.04$\pm$0.05 & -1.09$\pm$0.02 & -1.34$\pm$0.02\\
Lead/faint & -1.14$\pm$0.08 & -1.20$\pm$0.03 & -1.52$\pm$0.03\\
\hline
\alphaFf & & &\\
Trail & 0.09$\pm$0.01 & - & -\\
Lead & 0.17$\pm$0.01 & - & - \\
Lead/bright & 0.15$\pm$0.01 & - & -\\
Lead/faint & 0.21$\pm$0.02 & - & -\\
\hline
\hline           
\end{tabular}
\tablefoot{Each column corresponds to a different spectroscopic survey. We only use the stars with Prob(Sgr|$\Lam$,$\,\Bet$) > 0.2. The bright branch is defined as P(A|$\Lam$,$\,\Bet$)>0.2 while, for the faint, we use P(B|$\Lam$,$\,\Bet$)>0.2.}
\end{table}

Finally, we analysed the chemical composition of the Sgr stream. To do that, we used the spectroscopic metallicities of APOGEE, LAMOST and SEGUE. In the top panel of Fig.~\ref{fig:fig5} we show the \alphaF vs \FeH diagram of the stream using APOGEE (available for 1244 of our candidate sources), where we noted a slight bend in the sequence at a metallicity \FeH$\sim$\,-0.7, which is present throughout the whole stream. The sequence then remains flat until it bends again at FeH$\sim$-0.3. According to the recent work of \citet{Hasselquist2021}, who studied in detail the chemical composition of Sgr with 946 APOGEE stars, this is probably the result of a starburst that happened $\sim$5-7 Gyr ago, followed by a quenching in star formation some $\sim$3~Gyr ago, both most likely due to the influence of the MW. 

Below the \alphaF vs \FeH diagram we show, in three separate panels, the metallicity distribution function in the trailing, core and leading parts of the stream, from top to bottom respectively. 
All three surveys agree on the fact that the trailing tail is $\sim$0.3~dex more metal rich than the leading tails\footnote{We have checked that, for each individual catalogue, the apparent magnitude distribution of both arms is comparable. Thus, the \FeH difference between tails is unlikely to be an observational bias.}, as noted in previous works \citep[e.g.,][]{Yang2019,Hayes2020}. This is still the case even if we select stars with high Prob(Sgr|$\Lam$,$\Bet$) and high Prob($\varpi$)\footnote{This score helps us filter contamination based on values of {\tt parallax\_over\_error} too large. See Appendix~\ref{app:methods_poe}.}, and even if we refine our kinematic selection by using also the radial velocities shown in Fig.~\ref{fig:fig3}. One could argue that this metallicity difference is caused by the fact that the leading arm starts at $\Lam > 40^\circ$, thus corresponding to slightly older material than the trailing arm, but we have also check that the metallicity of the leading arm is systematically lower than the trailing arm at any $\Lam$. 
The same applies as a function of Galactic latitude, disfavouring a disc contamination bias. The \alphaF tells the same story given that the leading arm shows a higher \alphaF ratio. 
All in all, there is no doubt that the leading arm has a lower mean metallicity, as we report in Table~\ref{tab:metallicites}.

\begin{figure*}[t!]
   \centering
    \includegraphics[width=0.98\textwidth]{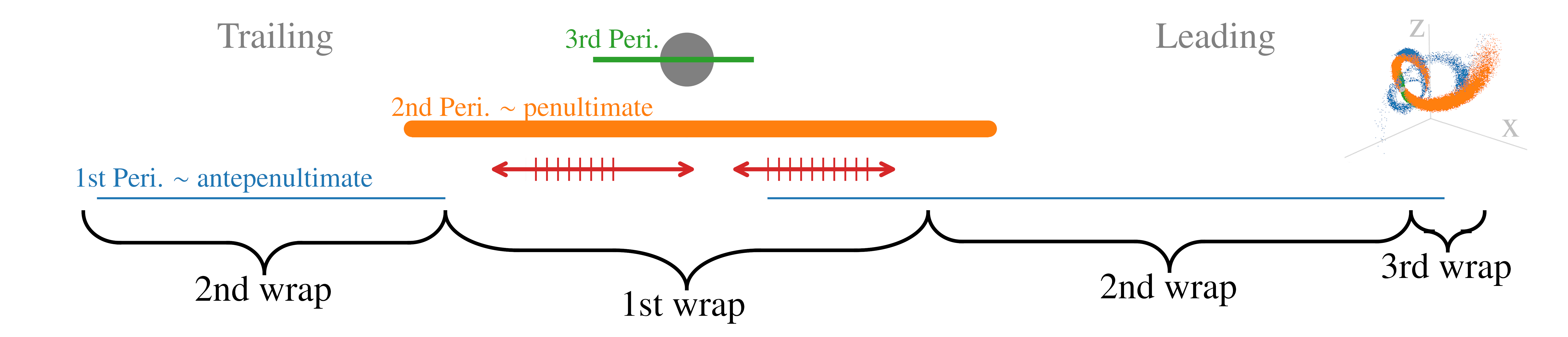}
    \caption{Schematic representation of the stream based on the \citetalias{Vasiliev2021} model. The lines represent the $\Lam$ extent of the material ejected during each perigalacticon, while their width is proportional to the amount of particles they contain. The blue line is for the stellar particles stripped near the first pericentre, the orange for the second one and the green for the most recent pericentre passage. We also include the nomenclature of the wraps into which we divide the stream: the first wrap is defined as $\Lam \in [-180^\circ, 180^\circ]$, the second as $\Lam \in [-540^\circ, -180^\circ] \cup [180^\circ, 540^\circ]$ and the third as $\Lam \geq 540^\circ$. As can be seen, the leading arm has stretched more than the trailing, probably due to the faster dynamical timescale. The 3D distribution in galactocentric coordinates of the stellar particles is shown in the top right inset. Finally, the red arrows represent the range of our data (see Fig.~\ref{fig:fig1}) and the hatched portion is where we detect the faint branches of the bifurcation.}
    \label{fig:fig6} 
\end{figure*}
\begin{figure*}[h!]
   \centering
    \includegraphics[width=0.98\textwidth]{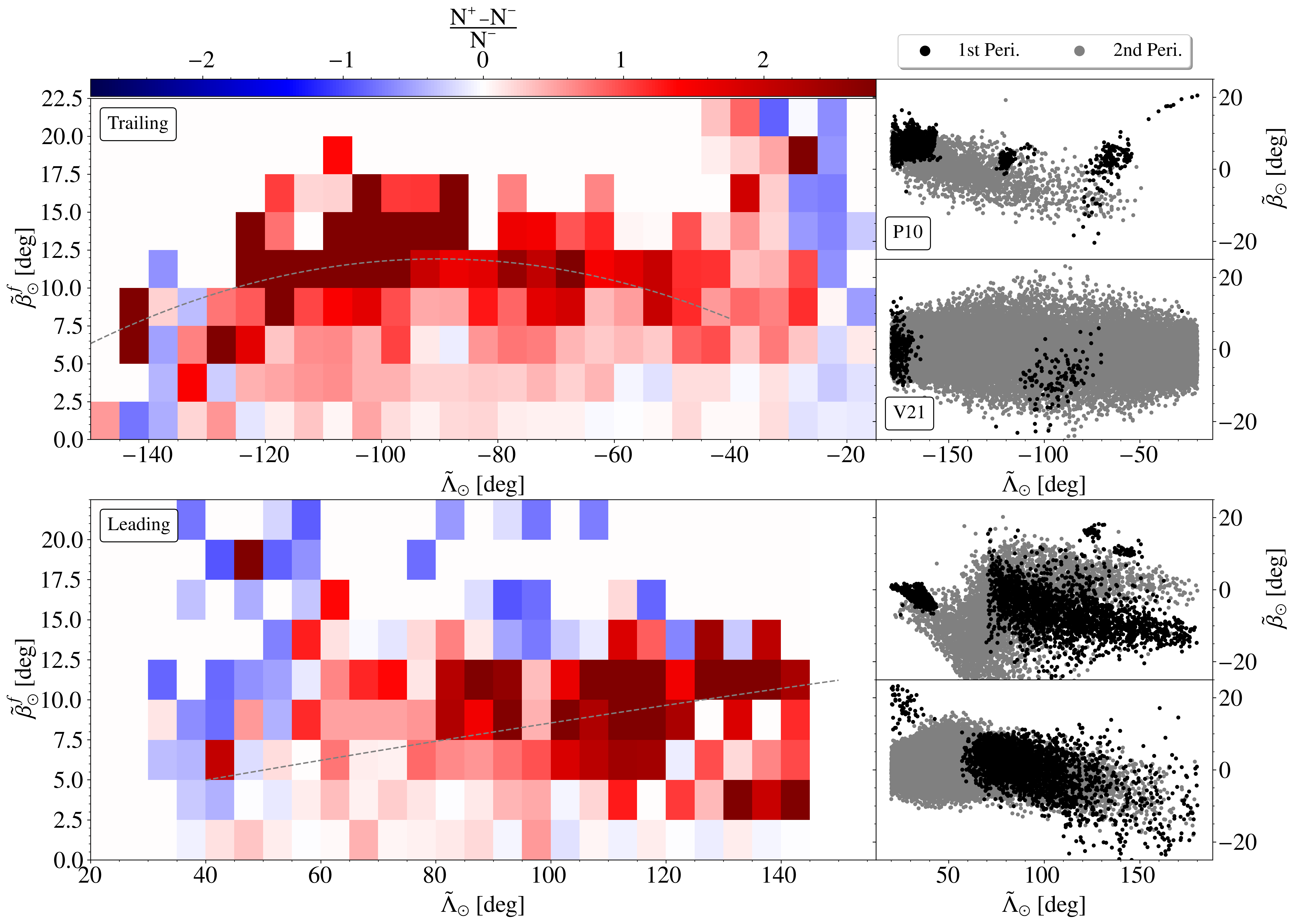}
    \caption{Normalised difference in counts between the two halves of the stream, folded in $\Bet$ along the line of highest density. The upper-script of $\tilde \beta^f_\odot$ stands for 'folded'. The grey dash lines correspond to the difference between the polynomials of the faint and bright branches (Table~\ref{tab:bifu_polys}). On the right, we include $\Lam$ vs $\Bet$ scatter plots of the stellar particles in the \citetalias{Vasiliev2021} and \citetalias{Penarrubia2010} N-body models, after applying a similar kinematic filter to the one used for the data, and separating by the time of stripping (black for the first pericentre, grey for the second). The bifurcation in the Sgr stream is obvious in the data and coincides with the functional form fitted in Sect.~\ref{sec:methods}. Also, the \citetalias{Penarrubia2010} model presents a clear bimodality, qualitatively similar to the data, produced by the first pericentre material.}
    \label{fig:fig7} 
\end{figure*}

Finally, we found that, in the leading arm, the faint branch is more metal poor than the bright branch. Also, the \alphaF for the faint is higher than for the bright, in accordance with the \FeHf. We could not do the same analysis in the trailing arm due to the footprints of the surveys used as there are too few stars on the faint branch. However, \citet{Koposov2012} shows that the faint branch of the trailing arm is also metal poor and probably made of ancient stripped material. Therefore, by association, it seems likely that this is also the case for the faint branch in the leading arm \citep[see also][]{Belokurov2014}. We note, however, that the mean \FeH of the bright branch in the leading arm is still lower than that of the trailing arm. The source of this difference is not clear but it could be due to the fact that in the leading arm there are older wraps of the stream mixed with the young leading arm, as we show in Fig.~\ref{fig:fig3}.

%--------------------------------------------------------------------
\section{Discussion}\label{sec:discussion}
Having studied the properties of our Sgr sample, we now focus on understanding the origin of the bifurcation. For that, we rely on two different models of the stream: \citetalias{Vasiliev2021}, which does not exhibit any bifurcation but is an excellent fit to the bright portion of the stream, and \citetalias{Penarrubia2010}, which reproduces the bifurcation in the leading arm by modelling the Sgr progenitor as a stellar disc rotating within a Dark matter halo.

In Fig.~\ref{fig:fig6} we show a schematic representation of the \citetalias{Vasiliev2021} model for the Sgr stream. Each colour represents an individual stripping event that the simulated dwarf galaxy has suffered. These correspond to, from bottom to top, the first, second and third pericentre passages respectively. The gap that exists between the beginning of the first pericentre stripped material and the progenitor (the grey sphere at the top centre) is caused by the quiescent time between the first and second pericentre where almost no material is stripped, and tells us about both the stripping history and orbit of the progenitor. We also include the notation that we use of the different wraps (see caption). For the remainder of the paper we refer to each portion of the stream with the format L1p1w, where the first letter distinguishes between (L)eading and (T)railing, the first number corresponds to the stripping time (1st, 2nd or 3rd pericentre), and the second one expresses the location along the stream (1st, 2nd or 3rd wrap).

With this schema in mind, we used Fig.~\ref{fig:fig7} to re-detect the bifurcation in the Sgr data and confirm that the morphology obtained from the 2D analysis of the stream (see Sect.~\ref{sec:methods}) is accurate. The left panels of Fig.~\ref{fig:fig7} are obtained after we fold the stream along the line in $\Bet$ of maximum density (i.e., using the equation in Table~\ref{tab:bifu_polys} for the bright branches) and subtract the stellar count on both sides of the stream. As a result, we obtained the two heat maps for both the trailing (top) and leading (bottom) arms. We note that the amount of stars in the faint branches represent roughly an excess of twice the number of expected stars at that position of the stream. \citet{Koposov2012} quantified, instead, the difference between the bright and faint branches in terms of integrated luminosity, finding that they differ by a factor of 5–10 in the trailing arm. 
We cannot repeat exactly the same exercise with the models due in part to the few particles available, but also because of the differences with the data. Instead, we show the distribution of the stellar particles in the $\Lam$ vs $\Bet$ space separating by the pericentre at which they got stripped: the first pericentre (black dots) or the second (grey dots). We have selected the particles in proper motion space along the corresponding kinematic tracks, thus mimicking our filter (Table~\ref{tab:parameters}) and allowing for a meaningful comparison with our data. 
We corroborate that \citetalias{Vasiliev2021} does not produce a bifurcation (the first and second pericentre material are well aligned in the first wrap) while, in contrast, the \citetalias{Penarrubia2010} model 'does' have a bifurcation in both leading and trailing tails, although the latter only extends up to $\Lam \lesssim -160^\circ$. We note also that, even with the proper motion selection, we still have some leading arm particles crossing the trailing arm at negative $\Lam$ in both models, which can be seen around $\Lam \sim -100^\circ$ in \citetalias{Vasiliev2021} and at $\Lam \sim -120^\circ$, $\Lam \sim -70^\circ$  in \citetalias{Penarrubia2010}. 

The utility of the schema presented in Fig.~\ref{fig:fig6} becomes apparent when trying to understand the origin of the bifurcation. As can be seen, exactly at the location of the observed bifurcation in the leading arm there is an overlap between L2p1w and L1p1w. This overlap is almost perfect in 6D phase-space according to the \citetalias{Vasiliev2021} model for the following reason: the particles that were stripped last on the first pericentre did not separate significantly from the progenitor until they approached together the next (second) pericentre. At that moment, the newly stripped material started to spread in the sky in almost the same way as the first pericentre material. Moreover, there are almost an order of magnitude less stripped particles at 1st pericentre compared to the 2nd pericentre according to \citetalias{Vasiliev2021}. All in all, if this is also the case for the real Sgr stream, it would be really hard to disentangle these two populations from our data with the level of precision we currently have. Indeed, we have been unable to detect two populations in phase-space along the bright branch of the leading arm but the uncertainties are far too large to discard that, within it, there is material from two different pericentres. The only observable that could aid us here would be their chemistry since the L1p1w contains particles that were less bound to Sgr (i.e., with smaller initial binding energies) than L2p1w, and this should correlate with their age and location within the progenitor which, in turn, should correlate with their \FeH and \alphaFf. As can be seen in Table~\ref{tab:metallicites}, we do find two chemical populations in the leading arm. However, these correspond to the bright (more metal rich) and faint (more metal poor) branches, which we interpret then as the L2p1w and L1p1w respectively. If that is the case, it means that the L2p1w and L1p1w material got ejected into slightly different orbits, causing a significant misalignment 'only' on the sky distribution. This is easily falsifiable for we should expect a sudden increase in metallicity at the $\Lam$ before which there is no L1p1w material ($\Lam \lesssim 65^\circ$ in the case of \citetalias{Vasiliev2021}). However, the spectroscopic surveys that we use, coming all from telescopes in the northern hemisphere, do not reach that part of the stream because it lies at negative declination.

While the \citetalias{Penarrubia2010} model has been shown to not reproduce well the observations (\citealt{Penarrubia2011}, and also our Fig.~\ref{fig:fig4}), it is a good example of how to produce a bifurcation in the Sgr stream. Surprisingly enough, its schema is qualitatively the same as that of \citetalias{Vasiliev2021}. However, the inclusion of a dominant rotational component causes three effects: first, it launches the L/T-1p material into a different orbital plane, causing a misalignment between L1p1w and L2p1w at the present day that resembles the observed bifurcation. Secondly, it ejects roughly three times more particles at its 1st pericentre than in the 2nd pericentre. This point is important because it means that, as expected, the relative brightness of the two branches is sensitive to the internal dynamics of the progenitor. Lastly, the resulting leading bifurcation has as the faint branch the L2p1w, while the L1p1w takes the role of the bright branch. This should cause the bright branch to be more metal poor, in contradiction with our observations. This last point, however, is not a critical issue since the configuration of the stream is sensitive to the relative angle between the angular momentum of the rotating component and the angular momentum of the Sgr orbit, which has never been explored systematically. So, there could in principle exist another configuration where the faint branch is in fact the L1p1w.

The trailing bifurcation is a bit more tricky to analyse in this context basically because of the huge gap between the progenitor and the 1st pericentre material seen in the models (Fig.~\ref{fig:fig6}). In other words, neither \citetalias{Vasiliev2021} nor \citetalias{Penarrubia2010} have a substantial T1p1w. Nonetheless, from the sky position of the T1p2w in \citetalias{Penarrubia2010} (see Fig.~\ref{fig:fig7}, upper-right panel) one can easily infer that, if it had material deposited along the T1p1w, this would indeed produce a bifurcation. All that is required to populate this part of the stream is that the stripping lasted longer after the first pericentre. The amount of stars that are stripped 'after' the pericentre passage depends (among other things) on the ratio between peri- and apocentric distance, which in turn is sensitive to dynamical friction. It is important to recall at this point that \citetalias{Vasiliev2021} does not have a live MW halo and, as a consequence, it relies on the Chandrasekhar description to account for dynamical friction, a recipe known to be inaccurate\footnote{We would like to stress that we are not implying that the \citetalias{Vasiliev2021} model is poorly constructed. It is the best Sgr stream model available. However, an inaccurate dynamical friction recipe can impact the stripping history.}. Therefore, it could be that by re-doing the simulation in a full N-body fashion, the stream presents a more prominent T1p1w. Another way to modify the stripping history is by altering the initial energy distribution of the stars prior to the first pericentre. This initial distribution function is not trivial to infer from the present-day observations. \citet{Lokas2010}, in particular, showed that it is actually possible to reproduce reasonably well the current properties of the remnant starting from a galaxy with a rotating disc that, as it falls inside the MW, gets stirred by the tidal field until is no longer recognisable.  Nonetheless, the fact that neither \citetalias{Vasiliev2021} or \citetalias{Penarrubia2010} have a T1p1w could just be a problem of sampling since both have very few particles and, in case of the former, only 4\% of the stellar particles are stripped at the first pericentre. We believe this last explanation could not account entirely for the missing component but is a factor to take into account.

Another interesting aspect of models \citetalias{Vasiliev2021} and \citetalias{Penarrubia2010} is that both predict different L1p2w and L1p3w. However, we have only been able to find, unequivocally, first pericentre material at the place where the two parts of the stream cross each other, as can be seen in Fig.~\ref{fig:fig3} (see also \citealt{Yang2019} and \citealt{Penarrubia2021}). This is the location with the least constraining power. We have tried actively looking for the rest of the L1p2w, specifically where it intersects the T2p1w, without success. If we were to find these stars they would constrain the whole ancient tail and, thus, allow us to better understand the formation of the bifurcation. We note, however, that the detection of \citet[][figures 5 and 11 in particular]{Navarrete2017} is naturally explained by this T2p1w-L1p2w crossing, to the extent that the point of maximum overlap at $\Lam \sim -70^\circ$ is accurately predicted by \citetalias{Vasiliev2021} model. Incidentally, we have a similar situation in the leading arm with the so called C branch \citep[e.g.,][]{Fellhauer2006}, which again is consistent with the crossing of, in this case, the L2p1w and the T1p2w \citep[comparing Table 2 of][with the V21 model]{Correnti2010}.

There have been other mechanisms proposed in the past to explain the bifurcations, most notably \citet{Fellhauer2006} where they tried to reproduce the faint branch by the overlaping of multiple old and young wraps, displaced relative to one another by the natural orbital precession introduced by the asphericity of the halo. However, their model did not match later observations of the stream. \citealt{Law2010}, on the other hand, discussed two other possibilities: orbital anisotropy within the progenitor and the influence of a Sgr satellite. 

Based on the current data and the discussion presented here, we conclude that the most likely scenario is that the antepenultimate pericentre material and the penultimate pericentre material have been ejected in slightly different orbital planes. Whether this difference comes from the rotation of the progenitor as proposed in \citetalias{Penarrubia2010} or, instead, from an unknown perturbation to the Sgr orbit, is still not clear. The former seems able to produce branches of comparable brightness since it alters the ejection rate at earlier times (compared to a non-rotating progenitor) whereas, in the latter, we might have to tweak the initial energy distribution of the Sgr stars to obtain the correct relative brightness between branches. This shall be the focus of our future studies on the stream.

%%%%%%%%%%%%%%%%%%%%%%%%%%%%%%%
\section{Conclusions}\label{sec:conclusions}

In this work, we have exploited the recent Gaia EDR3 astrometric sample to detect the Sgr stream and compile a list of more than 700k candidate stars. Thanks to the vast size of the sample obtained and the quality of the Gaia astrometry and photometry, we have been able to characterise the stream in great detail, specially the phase-space distribution of the trailing arm. As a result, we have quantified the correlation between the proper motions and the distance, which allows us to obtain a precise 6D picture of the stream. More importantly, we detect a significant over-density at $\Bet \geq 5^\circ$ throughout most part of the stream that we recognise as the bifurcation. Based on the data available, and thanks to the analysis of tailored N-body models of the stream, we conclude that the most likely origin for this feature is the orbital displacement of the material stripped shortly after the antepenultimate pericentre with respect to the material ejected during the penultimate pericentre.

With this work we have accomplished:

   \begin{enumerate}
      \item An update of our previous Sgr sample that is now $\sim3x$ larger,
      \item A numerical quantification of the sky distribution of the stream and the four branches (the two main ones, and the two responsible for the bifurcations),
      \item A precise characterisation of the kinematics (together with their correlation with distance) of the whole first wrap of the stream, and even beyond in the case of the radial velocity,
      \item The identification and parameterisation of the sharp edge in proper motion space that defines the physical envelope of the stream,
      \item The chemical properties of the different portions of the stream, showing that the lower metallicity of the leading arm is caused by the overlap with older wraps,
      \item A better understanding of the nature of the bifurcation as, according to our interpretation, it is due to the orbital misalignment of the material stripped $\gtrsim$2~Gyr ago with respect to the material stripped in the penultimate pericentre passage ($\sim$1~Gyr ago).
   \end{enumerate}

Looking forward, we should explore deeper the models of the stream to find under which circumstances a bifurcation could be formed. Especially challenging is the fact that both faint branches are on the same side of the bright branch which, in the scenario laid out in this work, would mean that a simple change in the orbital plane of the debris would not be enough as that would most certainly cause an X-shape in the sky. Anyhow, the streams produced by dwarf galaxies with a rotating component deserve more attention since there has never been a systematic exploration of the angle between the angular momentum of the rotating component and the orbital plane, the properties of said rotating component or its mass relative to the total mass of the dwarf galaxy, to mention the most relevant free parameters. Nonetheless, either in the case of a rotating progenitor or an external and unknown perturbation, there is scientific value in exploring these scenarios since both have implications on other fields. For instance, if the Sgr orbit was perturbed, we should be able to constrain the properties of said perturber and, perhaps, its impact also on the stellar distribution of the MW itself (if any).

To aid us in constraining the models, we should also obtain more observations of the stream, focusing on i) the chemical properties of the faint branches, specially for the trailing arm, and ii) locating the ancient wraps in other, more informative, places of the sky. By doing so, we could produce more accurate fits and understand better the interaction between Sgr, the MW and the Magellanic Clouds.

\begin{acknowledgements}
    We thank Raphaël Errani and Simon Rozier for all their useful comments. This work has made use of data from the European Space Agency (ESA) mission {\it Gaia} (\url{https://www.cosmos.esa.int/gaia}), processed by the {\it Gaia} Data Processing and Analysis Consortium (DPAC, \url{https://www.cosmos.esa.int/web/gaia/dpac/consortium}). Funding for the DPAC has been provided by national institutions, in particular the institutions participating in the {\it Gaia} Multilateral Agreement. 
    This work has been supported by the Agence Nationale de la Recherche (ANR project SEGAL ANR-19-CE31-0017). It has also received funding from the project ANR-18-CE31-0006 and from the European Research Council (ERC grant agreement No. 834148).
    TA acknowledges the grant RYC2018-025968-I funded by MCIN/AEI/10.13039/501100011033 and by ``ESF Investing in your future''. This work was (partially) funded by the Spanish MICIN/AEI/10.13039/501100011033 and by ``ERDF A way of making Europe'' by the ``European Union'' through grant RTI2018-095076-B-C21, and the Institute of Cosmos Sciences University of Barcelona (ICCUB, Unidad de Excelencia ’Mar\'{\i}a de Maeztu’) through grant CEX2019-000918-M. This project has received support from the DGAPA/UNAM PAPIIT program grant IG100319.
\end{acknowledgements}

\bibliographystyle{aa}
\bibliography{mybib}

\begin{appendix}

\section{Queries to the Gaia archive}\label{app:queries}

The results presented in this work have been obtained using the following queries to the \Gaia Archive:

\begin{lstlisting}[title=A,label={q1},caption=Query to obtain the proper motion histogram at a certain \HPf.]
SELECT COUNT(*) as N, pmra_index*BINSIZE as pmra, pmdec_index*BINSIZE as pmdec FROM (SELECT gaia.source_id, FLOOR(pmra/BINSIZE) AS pmra_index, FLOOR(pmdec/BINSIZE) AS pmdec_index, agn.catalogue_name FROM gaiaedr3.gaia_source AS gaia LEFT OUTER JOIN gaiaedr3.agn_cross_id AS agn ON agn.source_id = gaia.source_id WHERE gaia.source_id BETWEEN HPNUM*2**35*4**(12-LVL)  AND  (HPNUM+1)*2**35*4**(12-LVL) AND parallax_over_error BETWEEN -PLXCUT AND PLXCUT) AS sub WHERE sub.catalogue_name IS NULL GROUP BY pmra_index, pmdec_index
\end{lstlisting}

\noindent where \emph{HPNUM} is the number of the \HP tile being processed, \emph{LVL} is the level of the \HP tessellation (here, 6), \emph{BINSIZE} is the binning of the histogram in proper motion (0.12\,$\masyr$) and, finally, \emph{PLXCUT} is the maximum parallax quality allowed (see Eq.\,\ref{eq:plx_cut}).

\begin{lstlisting}[title=B,label={q2},caption=Query to obtain the main characteristics of the CMD for the stars in a peak in proper motion space.]
SELECT count(*) as N, ((SUM(phot_g_mean_mag * bp_rp)-(SUM(phot_g_mean_mag) * SUM(bp_rp)) / COUNT(*)))/(SQRT(SUM(phot_g_mean_mag * phot_g_mean_mag)-(SUM(phot_g_mean_mag)*SUM (phot_g_mean_mag)) / COUNT(*))*SQRT(SUM(bp_rp * bp_rp)-(SUM(bp_rp) * SUM(bp_rp)) / COUNT(*))) AS r, avg(phot_g_mean_mag) as g, avg(bp_rp) as colour, SQRT((COUNT(*)/(COUNT(*)-1))*(avg(POWER(bp_rp,2)) - POWER(avg(bp_rp),2))) as std_colour, avg(ra) as ra, avg(dec) as dec, avg(pmra) as pmra, avg(pmdec) as pmdec FROM gaiaedr3.gaia_source AS gaia LEFT OUTER JOIN gaiaedr3.agn_cross_id AS agn ON agn.source_id = gaia.source_id WHERE gaia.phot_g_mean_mag < maxG and gaia.g_rp is not Null AND gaia.source_id BETWEEN HPNUM*2**35*4**(12-LVL) AND (HPNUM+1)*2**35*4**(12-LVL) AND parallax_over_error BETWEEN -PLXCUT AND PLXCUT AND agn.catalogue_name IS NULL AND (POWER((gaia.pmra-PEAK_PMRA),2)+ POWER((gaia.pmdec-PEAK_PMRA),2)) < (2*PEAK_RADIUS)**2
\end{lstlisting}\label{query2}

\noindent from which we obtain the main photometric properties (mean, dispersion and Pearson's correlation coefficient) of an over-density in proper motion space. To do so, on one hand we limit the faintest magnitude allowed \emph{maxG} (19.75 mag in this work). On the other, we use the three parameters that defined any kinematic tree: \emph{PEAK\_PMRA}, \emph{PEAK\_PMDEC} and \emph{PEAK\_PMRADIUS}.

\section{Details of the methodology}\label{app:methods}

\subsection{Proper motion analysis}
Each proper motion histogram obtained at each \HP is transformed with the Wavelet Transformation \citep[WT,][]{Starck2002} to enhance and isolate the significant over-densities in kinematic space, which are then selected with the {\tt peak\_local\_max} algorithm included in the \textit{Python} package {\tt Scikit-image} \citep{Scikit-image}. The result at each \HP is a list of peaks for each scale of the WT characterised by a centre (pmra-pmdec coordinates) a radius, r = binsize$\,\times\,$2$^{\rm{scale}}$, and its WT coefficient defined by

\begin{equation}\label{Eq:wt}
    I(x,y) = c_N(x,y) + \sum_{j=0}^{N-1}\omega_j(x,y),
\end{equation}

\noindent where $I$ is the input histogram, $c_N$ is its smoothed version, and $\omega_j$ are the arrays of WT coefficients at each of the N scales $j\in[0,N-1]$.

Of course, any real structure will appear as an over-density at several (if not all) of the WT scales in which the original histogram is decomposed. However, in our previous works \citep{Antoja2020,Ramos2021} we chose, for simplicity, to select only the most prominent peak throughout all scales, one per \HPf. To be able to deal with all the information returned by the WT, we define the following two concepts.

\paragraph{Kinematic trees} We consider all peaks in all scales and we use a method to organised them in unique structures that we name \textit{kinematic trees}. It goes as follows: for each \HPf, we begin with a peak at scale 0 (the smallest layer), let us call it $p_i$, and then take all the peaks at scale 1, $p_{i+1}$, that fall within the circular area defined by $p_i$ ($r_i = binsize\,\times\,2^{i}$). We do not restrict ourselves to a single peak at the scale $i+1$, thus we are allowing these "trees" to "branch out". For each one of the peaks found at scale 1, we repeat the process but at scale 2. The process continues until we reach the largest scale, at which point we move to the next 0-scale peak and start over again. Once we have done this for every peak in our sample across all \HP we end up with a set of hierarchical structures that are characterised by the pmra-pmdec coordinates of the peak with the highest $\omega_j$, which in turns also sets the characteristic size of the whole structure, $r_j$. Each of these kinematic trees now represent a unique kinematic structure.

We note that, by construction, two branches can have a different seed (the 0-scale peak) and yet share the same higher level peak. In general, the peaks at the higher scales are big enough to cover a significant fraction of the 0-scale peaks. However, there must exist an interrupted link from scale 0 all the way up to the highest scale to link them together, which is not common.

\paragraph{Absorbing Poisson noise} We still have the problem that Poisson noise can create several small-scale peaks at the same \HP for a single object. In that case, they will quickly merge at the upper scales once the size of the peaks can absorb the noise. We correct for this "duplicity" by merging the trees that share the same scale-1 peak\footnote{If two trees have the same peak at layer $j$, all the peaks at upper layers $i \geq j$ will also be the same.} (r = 0.24\,mas\,yr$^{-1}$). In practice, this is very similar as starting from histograms with a coarser binning, but not exactly.

\paragraph{Parameters of the kinematic selection}
In Table~\ref{tab:parameters} we show the parameters used to isolate the kinematic signature of the Sgr stream (see Sect.~\ref{sec:methods}).

\begin{table*}
\caption{Parameters of Eq.\ref{eq:muVSlambda} used to select the proper motion structures belonging to the Sgr stream.}\label{tab:parameters}
\centering
\begin{tabular}{ccccc|cccc}
\hline\hline
     & \multicolumn{4}{c}{Trailing}  & \multicolumn{4}{c}{Leading} \\
     & \multicolumn{2}{c}{$\mu_{\Lam}$} & \multicolumn{2}{c}{$\mu_{\Bet}$} & \multicolumn{2}{c}{$\mu_{\Lam}$} & \multicolumn{2}{c}{$\mu_{\Bet}$}\\
      & U & L & U & L & U & L & U & L \\
\hline
$a_1$ & -1.1842     & -1.1842   & -1.2360   & -1.2360  & -1.1842 & -1.1842 & -1.2360 & -1.2360 \\
$a_2$ & 1.8000      & 1.5000    & -1.1800   & -1.4500  & 1.2000 & 1.2000 & -1.3200 & -1.4500 \\
$a_3$ & -0.1000     & -0.1000   & 0.16330   & 0.16330   & 0.1000 & 0.1000 & 0.2000 & 0.3500 \\
$a_4$ & 4.0000      & 1.4500    & -0.9800   & -2.8000   & 2.2807 & 1.5107 & -1.3000 & -2.5000 \\
$a_5$ & 8.0$\times10^{-3}$      & 8.0$\times10^{-3}$    & -7.3022$\times10^{-3}$& -7.3022$\times10^{-3}$   & 0.1606$\times10^{-3}$ & -0.0061$\times10^{-3}$ & -5.0$\times10^{-3}$ & -5.0$\times10^{-3}$ \\
$a_6$ & -0.9$\times10^{-5}$ & 0.9$\times10^{-5}$  & -4.0$\times10^{-5}$ & -4.0$\times10^{-5}$  & -0.5544$\times10^{-5}$ & -1.2544$\times10^{-5}$ & -0.05$\times10^{-5}$ & -0.05$\times10^{-5}$ \\
\hline
\hline           
\end{tabular}
\tablefoot{The selection is split into four parts, first separating between leading and trailing arms, then distinguishing between $\mu_{\Lam}$ and $\mu_{\Bet}$. Each of these four parts has slightly different parameters for the upper (U) and lower (L) bounds.}
\end{table*}

\paragraph{K-means clustering}
After selecting the kinematic trees based on the proper motion trends of the Sgr stream, we obtained the photometric summary statistics of each of them using query~\ref{q2}. We then ran a k-means clustering with 6 clusters on the vector ($\Lam$, $\Bet$, $r_{G-colour}$, $<G>$, $<G_{BP-RP}>$, $\sigma_{G_{BP-RP}}$), which we re-normalised so that all quantities have a comparable range.

\begin{figure}[]\label{fig:figB1} 
   \centering
    \includegraphics[width=0.49\textwidth]{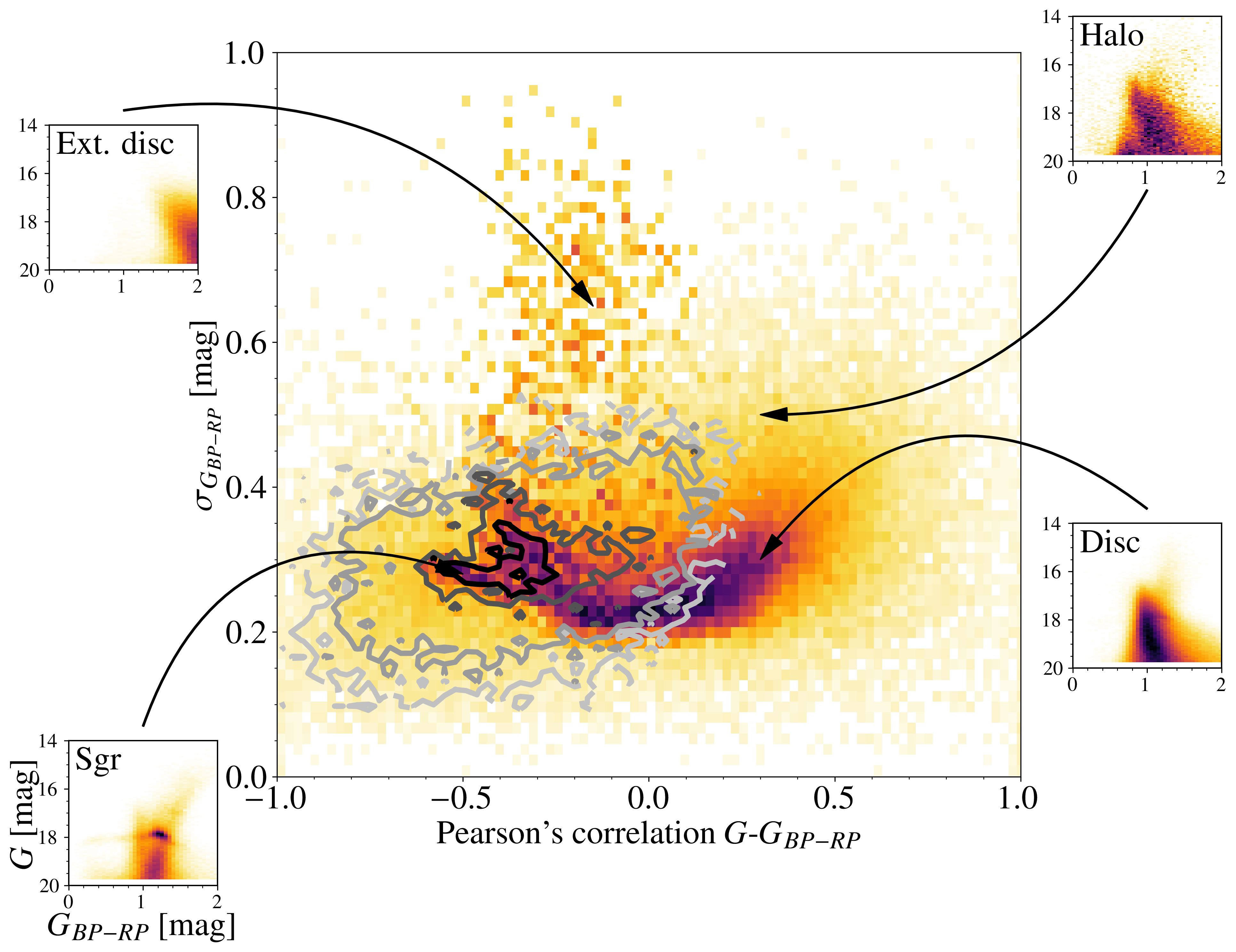}
    \caption{Result of the K-means clustering. Centre: histogram in the space of $r_{G-colour}$ vs $\sigma_{G_{BP-RP}}$ weighted by number of stars of the kinematic trees that fulfil the filter of Table~\ref{tab:parameters}. The contours correspond to the kinematic trees labelled as Sgr stream. Bottom left: example of the CMD obtained from the kinematic trees associated with Sgr. Bottom right: same but kinematic trees associated to the disc. Top right: same but for the halo. Top left: same but for the very extincted disc.}
\end{figure}

The results can be seen in Fig.~\ref{fig:figB1}, where we show all kinematic trees in the space of $r_{G-colour}$ against $\sigma_{G_{BP-RP}}$, highlighting those tagged as Sgr stream (dark contours). Surrounding the main panel, we included examples of the CMDs that the kinematic trees contain depending on their photometric properties, showing that the K-means clustering has indeed separated the different populations correctly.

\subsection{Using parallax\_over\_error to discern contamination}\label{app:methods_poe}

When looking at the CMDs of our final sample at different position along the stream, we noticed a persistent source of contamination that does not follow any isochrone shape. Upon further inspection, we see that these correspond to either halo or thick disc stars (depending on the region on the sky). Luckily, the {\tt parallax\_over\_error} distribution of these contaminant sources does not follow the same distribution as the Sagittarius stars. In fact, the non-Sgr stars tend to have larger values of {\tt parallax\_over\_error}, thus deforming an otherwise symmetric Gaussian distribution. We took advantage of this by:

\begin{enumerate}
    \item Finding the mode of the distribution.
    \item Mirroring only the part of the histogram left to said mode (the negative side). 
    \item Stitching together the original negative side with its mirrored version.
    \item Fitting a normal distribution to it.
\end{enumerate}

This results in a symmetric probability distribution of {\tt parallax\_over\_error} that describes accurately the Sgr stream ($\mu = 0.29\,$mas, $\sigma = 1.12\,$mas). In reality, one would have to do this at 'every' position in the sky both because of the distance gradient and because the parallax zero-point is position dependent \citep{Lindegren2021}. In practice, the Sgr coordinates are almost aligned with the ICRS frame, causing the stream to do only small excursions in ecliptic latitude and, as a consequence, the variations on the zero-point are very small. On the other hand, the distance gradient is almost negligible once inverted into parallaxes, especially because the progenitor contains many more stars than the tidal tails. 

Finally, we normalised the p.d.f. and assigned a probability to each star based on how close the observed distribution is to the expected number. In other words, the stars that lie close to the mode will get a score close to 1, while those lying in a region where there is a lot of contamination will get a low score. We call this score Prob($\varpi$).

\subsection{Adding radial velocities}\label{app:methods_vlos}
The radial velocities used in this work have been compiled from 5 different catalogues: APOGEE, LAMOST, SEGUE, Gaia DR2 and SIMBAD. Each catalogue has its own caveats but we decided to simplify the merging by computing the median radial velocities with just the measurements that pass the following filters:

\begin{itemize}
    \item APOGEE: signal to noise larger than 10,
    \item LAMOST: error in velocity smaller than 20 km$\,$s$^{-1}$,
    \item SEGUE: based on the recommendations described in their web page, we used the cut {\tt (sp.zwarning = 0 OR sp.zwarning = 16)
AND sp.elodiervfinalerr != 0
AND sp.snr > 35},
    \item Gaia: no cut,
    \item SIMBAD: no cut.
\end{itemize}

\section{Sagittarius EDR3 sample}\label{app:sample}
In table \ref{tab:sample} we show 10 stars of our final sample of the Sgr stream, which can be found online at CDS. The description of the columns can be seen in the caption of the table.

\begin{table*}\label{tab:sample}
\caption{Ten rows of our Sgr sample.}
    \footnotesize

    \setlength\tabcolsep{3pt}
    {
\begin{center}
\begin{tabular}{ccccccccccc}
\hline\hline
source\_id &  $G$ & $G_{BP-RP}$ & ra & dec &     
$\mu_{\alpha*}$&  $\mu_{\delta}$ & $\sigma_{\mu_{\alpha*}}$ &  $\sigma_{\mu_{\delta}}$ & D$_{WGBPRP}$ & $\sigma_{D_{WGBPRP}}$  \\

& mag & mag & [$^\circ$] & [$^\circ$] &
[mas\,yr$^{-1}$] &  [mas\,yr$^{-1}$]  & [mas\,yr$^{-1}$]  &  [mas\,yr$^{-1}$] & [kpc] & [kpc]\\
\hline
6754363563665328512 & 18.724 & 1.183 & 296.195 & -27.978 & -2.671 & -1.914 & 0.277 & 0.197 &  &   \\
6744103853167673344 & 17.403 & 1.05 & 289.404 & -33.232 & -2.545 & -1.278 & 0.1 & 0.086 &  &   \\
6816189931422835328 & 17.162 & 0.644 & 323.496 & -22.274 & -2.648 & -3.075 & 0.094 & 0.055 & 21.553 & 1.863  \\
6761080621962994944 & 14.892 & 2.131 & 283.133 & -30.639 & -2.667 & -1.455 & 0.032 & 0.025 &  &   \\
2501913798194038784 & 16.156 & 1.427 & 39.755 & 0.904 & -0.148 & -2.139 & 0.054 & 0.051 &  &   \\
29057038601187712 & 17.178 & 1.471 & 44.713 & 13.416 & 0.022 & -1.835 & 0.099 & 0.089 &  &   \\
4415033523372538240 & 19.409 & 0.316 & 228.157 & -2.283 & -1.553 & -0.136 & 0.401 & 0.287 &  &   \\
6759908130238391040 & 18.2 & 1.239 & 284.039 & -32.635 & -2.683 & -0.939 & 0.202 & 0.192 &  &   \\
3267035009066699776 & 15.144 & 2.221 & 47.957 & 1.254 & -0.024 & -1.757 & 0.032 & 0.028 &  &   \\
6324577220124659968 & 12.565 & 0.945 & 217.402 & -11.771 & -0.804 & -0.921 & 0.279 & 0.22 &  &   \\

\hline
\hline
\end{tabular}
\end{center}

\begin{center}
\begin{tabular}{cccccc}
\hline\hline
  v$_{los}$ & Source v$_{los}$ &  [Mg/H] APOGEE &  \FeH APOGEE  & \FeH LAMOST  & \FeH SEGUE \\

 [\kms{}] & & [dex] & [dex] & [dex] & [dex] \\
\hline
&&&& \\
&&&& \\
&&&& \\
134.223&A&-0.584&-0.547& \\
-144.927&SC&&&&-0.51 \\
-185.02&L&&&-0.845 \\
&&&& \\
&&&& \\
-126.318&ASC&-1.177&-1.063&&-9999. \\
-4.228&GC&&& \\
\hline
\hline
\end{tabular}
\end{center}

\begin{center}
\begin{tabular}{cccc}
\hline\hline
  Prob(A | $\Lam$, $\Bet$) & Prob(B | $\Lam$, $\Bet$) & Prob(Sgr | $\Lam$, $\Bet$) & Prob($\varpi$)\\

& & & \\
\hline
0.196&0.593&0.673&0.96 \\
0.877&0.0&0.877&0.98 \\
0.002&0.806&0.806&0.97 \\
0.983&0.0&0.983&1.00 \\
0.999&0.0&0.999&0.9 \\
0.003&0.786&0.787&0.97 \\
0.081&0.976&0.977&0.98 \\
0.862&0.0&0.862&0.43 \\
0.335&0.0&0.335&0.99 \\
0.01&0.0&0.01&0.64 \\

\hline
\hline
\end{tabular}
\end{center}

}
\tablefoot{The first columns contains the source id of the star followed, in columns 2 and 3, by the apparent $G$ magnitude and the BP-RP colour. Then, columns 4 to 9 contain, respectively, the right ascension, declination, proper motion in both, and their uncertainties. Then, on the tenth and eleventh columns we include the Period-Wesenheit distances and their uncertainties for our sub-sample of RR Lyrae. Columns 12 and 13 correspond, respectively, to the value of radial velocity adopted and the catalogues used to obtain it (A - APOGEE, L - LAMOST, S - SEGUE, G - Gaia, C - Simbad). Then, on column 14, we show the magnesium abundance from APOGEE. The next three columns show the metallicites from APOGEE, LAMOST and SEGUE, in that order. Finally, the last four columns carry the different probabilities that we have calculated: probability of a star of belonging to the bright branch (col.~18), of belonging to the faint (col.~19), of belonging to the Sgr stream (col.~20) and the {\tt parallax\_over\_error} score (col.~21).
%The full table is available at the \href{http://cdsarc.u-strasbg.fr/viz-bin/cat/J/A+A}{CDS}.
}
\end{table*}

\end{appendix}

\end{document}